\documentclass[%
 aip,
 amsmath,amssymb,
 reprint,%
]{revtex4-1}

\usepackage{graphicx}
\usepackage{dcolumn}
\usepackage{bm}

\usepackage[utf8]{inputenc}
\usepackage[T1]{fontenc}
\usepackage{mathptmx}
\usepackage{etoolbox}

\makeatletter
\def\@email#1#2{%
 \endgroup
 \patchcmd{\titleblock@produce}
  {\frontmatter@RRAPformat}
  {\frontmatter@RRAPformat{\produce@RRAP{*#1\href{mailto:#2}{#2}}}\frontmatter@RRAPformat}
  {}{}
}%
\makeatother
\begin{document}

\preprint{AIP/123-QED}

\title[Bound modes in the continuum based phononic waveguides]{Bound modes in the continuum based phononic waveguides}
\author{Adib Rahman}
 
\author{Raj Kumar Pal}%
\email{rkpal@ksu.edu}
\affiliation{ Department of Mechanical and Nuclear Engineering, Kansas State University, Manhattan, Kansas 66506, USA}

\date{\today}

\begin{abstract}
We analytically predict and numerically demonstrate the existence of a family of bound modes in the continuum (BICs) in bi-layered spring mass chains. A coupled array of such chains is then used to illustrate transversely bound waves propagating along a channel in a lattice. We start by considering the compact region formed by coupling two spring mass chains with defects and predict bound modes arising due to reflection symmetries in this region. Dispersion analysis of a waveguide consisting of an array of appropriately coupled bi-layered chains reveals the presence of a branch having bound modes in the passband. Finally, detailed numerical analyses verify the existence of a BIC and its propagation through the waveguide at passband frequencies without energy leakage. The framework allows to achieve BICs and their propagation, for any arbitrary size and location of the compact region. Such BICs open avenues for novel classes of resonators with extremely high $Q$ factors due to zero energy leakage and allow for guiding confined waves in structures without requiring bandgaps. 
\end{abstract}

\maketitle

\section{Introduction}

Bound modes in the continuum (BICs) are a unique class of localized modes satisfying two properties. The first is that their frequency lies in the continuous spectrum of propagating waves, i.e., in the pass band and the second is that the wave amplitude is zero outside a compact spatial region~\cite{hsu2016bound}. In contrast, conventional bound modes lie in bandgaps, while localized modes in pass bands are typically leaky, with the wave amplitude gradually decreasing with distance from the wave center.  BICs were first predicted in 1929 by von Neumann and Wigner in the context of quantum mechanics~\cite{neumann1993merkwurdige}. Their idea was based on a complex artificial potential that was hard to achieve in real materials, so the idea did not get attention for several decades. The concept of BICs was revived again when their properties were explored in atomic systems~\cite{FONDA1963123,PhysRevA.11.446} and possibility of achieving them in other physical domains was revealed. BICs were first observed experimentally in acoustics by ``Wake Shedding Experiment" in 1966~\cite{parker1966resonance}. Today, BICs  are an active research field in multiple fields, particularly because their zero leakage property allows the possibility to achieve modes with very high quality factors ($Q$ factor) \cite{KOSHELEV2019836}. Potential applications of BICs include lasing \cite{imada1999coherent,hirose2014watt,kodigala2017lasing}, filtering \cite{Doskolovich:19,foley2014symmetry}, sensing \cite{Romano:19}, supersonic surface acoustic devices \cite{kawachi2001optimal,naumenko2003surface}, and guiding energy \cite{benabid2002stimulated,couny2007generation}.    

In the last few years, advances in manufacturing have revived interest in BICs, particularly using photonic metamaterials, as a route to achieve resonators with infinite $Q$ factors. Examples of BIC supporting structures include photonic crystal slabs~\cite{PhysRevB.61.2090,PhysRevLett.113.037401,PhysRevLett.113.257401}, waveguide array \cite{PhysRevLett.111.240403}, and metasurfaces~\cite{PhysRevLett.121.193903,Cai:21}. Most BICs fall into two categories: symmetry protected and accidental. Symmetry protected BICs arise when spatial symmetry of a localized or defect mode is 
incompatible with the symmetry of the propagating modes. They have been observed experimentally in a dielectric slab with square array of cylinders \cite{hsu2013observation}, a periodic chain of dielectric disks~\cite{sadrieva2019experimental} as well as in an optical waveguide~\cite{plotnik2011experimental}. Accidental BICs arise due to cancellation of coupling coefficients to the radiating or propagating waves by carefully tuning one or several system parameters. An example of this category is the Fabry-Parot BIC, where a BIC is formed by destructive interference of waves reflected from a large distance~\cite{PhysRevLett.100.183902}. Structures supporting quasi-BICs \cite{taghizadeh2017quasi,abujetas2021high} with giant $Q$ factors are emerging as an alternative to BICs, as the number of structures supporting BICs is limited. 

Compared to photonics, BICs have been much less investigated in elastic media. The main challenge in achieving elastic BICs is the simultaneous presence of longitudinal and transverse waves with distinct dispersion relations. BICs require modes that do not couple or hybridize with either of these propagating wave types. Recently, Haq and Shabanov~\cite{haq2021bound} theoretically predicted an elastic BIC containing coupled in-plane transverse and longitudinal waves in a system of two periodic arrays of cylinders. Other examples include the demonstration of three distinct in-plane mechanical BICs in micro-scale slab-on-substrate phononic crystal \cite{tong2020observation} and the existence of quasi-BICs in a semi-infinite plate attached to a periodic waveguide \cite{cao2021elastic}. Cao \textit{et al.}\cite{cao2021elastic} observed the quasi-BIC in an equivalent finite structure and discussed how to turn the quasi BICs to BICs by exploiting Fabry-Perot resonance frequency. There have also been works showing mechanical BICs coupled to optical resonance in various opto-mechanical systems \cite{Zhao:19,chen2016mechanical}. All these prior works in elastic media focus on achieving BICs in specific structural configurations, such as Fabry-Perot resonators in periodic cylinders separated by large distance~\cite{haq2021bound} or perfect mode conversion in a plate attach to waveguide \cite{cao2021elastic}. A general framework or design paradigm that can enable to achieve BICs in various structures such as beams, plates, shells and solids, translating across length scales and material properties, would be of interest. In addition, the coupling of an array of BICs remain unexplored and offer opportunities for leakage free wave propagation along waveguides.

\begin{figure}[!t]
    \centering
    \includegraphics[height = 3.5cm]{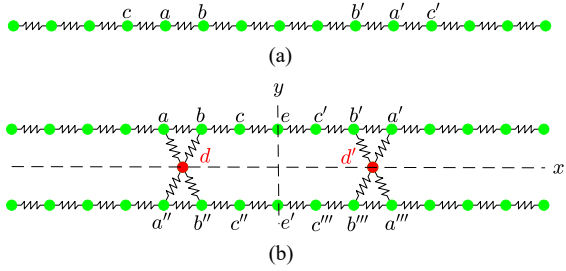}
     \caption{(a) Chain comprising of point masses (green nodes) connected by springs. It is challenging to achieve BICs by modifying mass or stiffness in this chain. (b) Proposed concept: two identical chains coupled with two defects (red nodes) create a symmetric compact region where bound modes arise.}
    \label{fig:chain}
\end{figure}

This work presents a framework to induce BICs in an arbitrary region of a one-dimensional ($1D$) periodic meta-structure. The meta-structure considered is represented by discrete springs and masses (schematic in Fig.~\ref{fig:chain}a) and BICs are induced by adding defects in a compact region (Fig.~\ref{fig:chain}b). We present a procedure to achieve a family of BIC modes by exploiting reflection symmetries in the compact region. Finally, a waveguide is constructed using an array of coupled spring-mass chains supporting BICs. This waveguide supports wave propagation without any energy leakage into the surrounding structure along the transverse direction. The paper is organized as follows: Sec.~\ref{sec:bicChain} presents the proposed design, derives the mode shapes of BICs and the dispersion curves of the waveguide.  Numerical simulations verifying the concept are presented in Sec.~\ref{sec:numer} and findings are summarized in Sec.~\ref{sec:conc}.

\section{Theory: BIC mode shapes and waveguides}\label{sec:bicChain}

We first discuss the proposed concept of inducing BICs by adding defects coupling 2 spring-mass chains. Then, symmetry conditions are used to derive the resulting mode shapes and verify if they are BICs. Next, dispersion analysis of a periodic structure comprising of an array of defect embedded spring-mass chains is presented. This analysis shows the presence of a dispersion branch supporting the propagation of transversely bound modes. This dispersion branch is used in the next section to illustrate leakage free wave transport along a waveguide. 

\subsection{BICs coupled spring mass chain}
Let us consider the problem of achieving BICs in a compact region of a discrete spring mass chain, whose schematic is illustrated in Fig.~\ref{fig:chain}a. Each node has a point mass $m$ with one degree of freedom and can move out of plane. The springs are identical and linear with stiffness $k$. The physical system we have in mind is a string in tension with rigid spheres attached to it at periodic intervals, similar to the setup considered in ~\cite{manktelow2014analysis}. We choose this spring-mass chain as the candidate for achieving BICs since it is the simplest structure whose dispersion behavior has a passband and stopband in the frequency-wavenumber domain~\cite{hussein2014dynamics}. 

To achieve BICs in a compact region of the chain, say between masses $a$ and $a'$ in Fig.~\ref{fig:chain}a, these two masses have to be at rest so that they do not induce any displacement outside the region. In addition, the masses located between them should vibrate at a frequency lying in the passband. A simple strategy is to modify the masses and springs in the compact region that leads to a localized defect mode and keeps the masses $a$ and $a^\prime$ at rest. Let us see why this approach cannot lead to BICs. To have a BIC between $a$ and $a^\prime$, we require that masses $a$ and $c$ are at rest, while mass $b$ has non-zero displacement $u_b$. Such a mode cannot satisfy the force equilibrium of mass $a$ since it is subject to a force $k u_b$. Hence mass $b$ also has to be at rest. Repeating the above argument for mass $b$, we see that if masses $a$, $c$, $a^\prime$ and $c^\prime$ are at rest as required for a BIC, then all masses between them are also at rest due to force equilibrium condition. 

To address this obstacle, let us consider two identical spring mass chains and couple them with two defect masses (red nodes in Fig.~\ref{fig:chain}b). The defect masses ($m_d$) are different from the masses in a spring mass chain (green nodes), while the stiffness of springs connecting them to chains are identical and equal to $k$. The two defects create a reflection symmetric compact region about $x$ and $y$ axes as illustrated in Fig.~\ref{fig:chain}b. Note that the whole structure need not have reflection symmetry about the $y$ axis as the location of the compact region can be completely arbitrary along the structure. 

Let us discuss the key idea that leads to BICs in the compact region between the two defect masses. Note that if the net force acting on mass $a$ from springs connecting masses $b$ and $d$ are zero, then mass $a$ will be at rest. In this case, all the masses to its left will also be at rest. The net force on mass $a$ will be zero when masses $b$ and $d$ have equal and opposite displacements. Applying a similar argument to masses $b'$, $b''$, $b'''$ and their corresponding defects, we see that masses $a'$, $a''$ and $a'''$ can be at rest and the masses outside the compact region are also at rest. Hence if we have a mode shape satisfying the requirement of the defect mass $d$ moving in opposite directions to 
mass $b$, then BICs can arise in the compact region. 

In this work, we will primarily illustrate BICs in a  compact region having 5 masses of each spring mass chain. However, the procedure we introduce here allows us to achieve BICs for any number of masses in a compact region embedded in a longer uniform spring mass chain. We also remark here that this idea of cancelling the force on mass $a$ can also be implemented on a single spring mass chain. Indeed, connecting a defect mass $d$ with springs to masses $a$ and $b$ will give the same effect. Since our motivating physical setup of a string in tension does not allow for hanging masses, we chose to implement the idea using two spring mass chains. 

\subsection{Bound mode shapes}
Let us now determine the frequencies and mode shapes of the bound modes in the compact region in Fig.~\ref{fig:chain}b. The governing equation of motion of a mass $p$ subjected to an external force $F_p$ is $m\ddot{u}_p + \sum_{\langle p,q\rangle} k(u_p-u_q) = F_p$. Here, $\langle p , q \rangle$ denotes a spring connecting mass $p$ and $q$. Let $l$ be the distance between two adjacent masses in the chain and let $\omega_0 = \sqrt{k/m}$ be the reference frequency. We express the physical quantities non-dimensionally by normalizing the frequency as $\Omega  = \omega/\omega_0$, time as $\tau = {\omega_0} t$, defect mass as $\alpha = m_d/m$ and force as $f_p =  F_p/kl$. Using these quantities, the equation of motion for the masses in the two main chains (green nodes in Fig.~\ref{fig:chain}b) becomes 
\begin{equation}
    \frac{d^2u_p}{d\tau^2} + \sum_{\langle p,q \rangle}(u_p-u_q) = f_p.
    \label{eq:non_dim_gov}
\end{equation}
The governing equation for the defect is $\alpha \ddot{u}_d + \sum_{\langle d,q\rangle} (u_d-u_q) = f_d$. The resulting system of equations can be expressed in matrix form as $\bm{M}\ddot{\bm{u}}+ \bm{K}\bm{u} = \bm{f}$, where $\bm{M}$ and $\bm{K}$ are the mass and stiffness matrices, respectively. Now, to determine the natural frequencies as well as corresponding mode shapes, we set $\bm{f} = 0$ and seek solutions of the form $\bm{u}(\tau) = \bm{u}e^{i\Omega \tau}$. The governing equations give an eigenvalue problem of the form $\bm{Ku} = \Omega^2\bm{Mu}$. Here  $\Omega$ is the non-dimensional natural frequency and $\bm{u}$ is the corresponding mode shape. 

We exploit reflection symmetries in the compact region to determine the mode shapes. There is a close relation between a spatial symmetry in a structure and its mode shapes~\cite{dresselhaus2007group}. In general, any spatial transformation (like reflection, rotation, translation) can be expressed as an operator. If a structure has a symmetry, then its mode shapes are also eigenvectors of the symmetry operator. The corresponding eigenvalues yield key information about the symmetry of the respective mode shapes. For example, let us consider a structure that has reflection symmetry about an axis. The reflection symmetry operator $\bm{R}$ has two eigenvalues: $\pm 1$. Thus, for each mode shape $\bm{v}$ in this structure, we will have $\bm{R}\bm{v} = +\bm{v}$ or $\bm{Rv} = -\bm{v}$. The mode shapes satisfying $\bm{Rv}=\bm{v}$ are symmetric (even) about the reflection symmetry axis, while those satisfying $\bm{Rv}=-\bm{v}$ are anti-symmetric (odd) about the symmetry axis. Hence reflection symmetry lead to two different classes of mode shapes in the structure. 
    
Let us apply the above concept to see the type of bound mode shapes that can exist in the compact region in Fig.~\ref{fig:chain}b. It has reflection symmetry about the $x$ and $y$ axes and thus we have two symmetry operators, $\bm{R}_x$ and $\bm{R}_y$. Thus each bound mode is either symmetric or anti-symmetric about both axes, in addition to having zero displacement outside the compact region. These symmetry conditions give 4 types of mode shapes satisfying
\begin{itemize}
    \item $u(x,y)=u(-x,y)=u(x,-y)$: even about both $x$ and $y$ axes
    \item $u(x,y)=u(-x,y)=-u(x,-y)$: even about $x$ axis but odd about $y$ axis
    \item $u(x,y)=-u(-x,y)=u(x,-y)$: odd about $x$ axis but even about $y$ axis
    \item $u(x,y)=-u(-x,y)=-u(x,-y)$: odd about both $x$ and $y$ axes
\end{itemize}
The symmetry conditions relating the displacement $u_p$ of mass $p$ is expressed in the form $u(x,y)$ above for the convenience of the reader. For example, $u_c = u(-1,1)$ and $u_{c^\prime} = u(1,1)$, following the notation in Fig.~\ref{fig:chain}b. 

In the above set of mode shapes, bound modes can only arise when the displacement is symmetric (even) about the $x$ axis. Let us see why bound modes that are anti-symmetric (odd) about $x$-axis are not possible. In such modes, $u_d = 0$ and a bound mode requires that $u_a = 0$. Equilibrium of mass $a$ requires that $u_b=0$ and thus the displacement of all masses in the compact region is zero, i.e., not a valid mode shape. 
Hence, there can be two types of BICs in the compact structure. The first type is even about both $x$ and $y$ axes and the second is even about the $x$ axis but odd about the $y$ axis, or in short, even and odd BICs about the $y$ axis.

Let us now determine the mode shape and frequency of the bound modes, followed by the defect mass required to support them. Recall that in bound modes, the displacement is zero outside the compact region and so it suffices to consider the masses in this region. Let us first consider even BIC mode shapes. We apply the following conditions on the mode shape to reduce the number of unknown displacements: 1. symmetry about $x$ axis implies $u_b=u_{b^{\prime\prime}}$, $u_c = u_{c^{\prime\prime}}$ and $u_e = u_{e'}$,  2. symmetry about $y$-axis implies $u_b = u_{b^\prime}$, $u_{b^{\prime\prime}} = u_{b^{\prime\prime\prime}}$ and similar constraint for masses $c$ and $d$, 3. equal and opposite displacement of masses $b$ and $d$ implies $u_b = -u_d$ and 4. displacements of masses $a$, $a'$, $a''$ and $a'''$ are zero. 
Using these constraints, the only unknown displacements are those of masses $b$, $c$, and $e$. The mass and stiffness matrices for even bound modes, $\bm{M}_{e}$ and $\bm{K}_{e}$, corresponding to the displacement vector $\bm{u}_e = (u_b, u_c, u_e)^T$ then become
\begin{equation*}
    \bm{M}_{e} = \begin{bmatrix}
    1 &0 &0\\
    0 &1 &0\\
    0 &0 &1
    \end{bmatrix} , \quad \bm{K}_{e} = \begin{bmatrix}
    4 &-1 &0\\
    -1 &2 &-1\\
    0 &-2 &2
    \end{bmatrix}.
\end{equation*}
The resulting eigenvalue problem is $\bm{K}_e \bm{u}_e = \Omega^2 \bm{M}\bm{u}_e$ and its solution gives the 
non-dimensional bound mode frequencies $\Omega=0.66, 1.73, 2.14$. 
These are the frequencies of even bound modes in the compact region illustrated in Fig.~\ref{fig:chain}b. 

Next, let us determine the defect mass $\alpha$ that is required to support each of these bound modes. Recall that we set $u_d = -u_b$ to derive the above frequencies and we still need to satisfy the governing equation for the defect mass. To this end, let us first determine the effective stiffness of the springs connected to $d$. Since mass $a$ is stationary, the effective stiffness due to spring $\langle d ,a\rangle$ is $k$. Similarly, since the mass $b$ has equal and opposite displacement to mass $d$, we have $k(u_d-u_b)=2k u_d$ and thus the effective stiffness due to spring $\langle d,b\rangle$ is $2k$. Similar arguments can be applied to springs $\langle d , a^\prime\rangle $ and $\langle d, b^\prime\rangle$. Thus the total equivalent stiffness of springs connected to mass $d$ is $2(k+2k)=6k$. The non-dimensional governing equation for $d$ thus becomes $(-\alpha \Omega^2 + 6)u_d =0$. Since a bound mode has non-zero $u_d$, we have $\alpha = 6/\Omega^2$. Using this expression, the non-dimensional defect masses $\alpha$ are 13.68, 2, and 1.32, corresponding to frequencies 0.66, 1.73, and 2.14, respectively.  

\begin{figure}[!t]
    \centering
    \includegraphics[height=9cm]{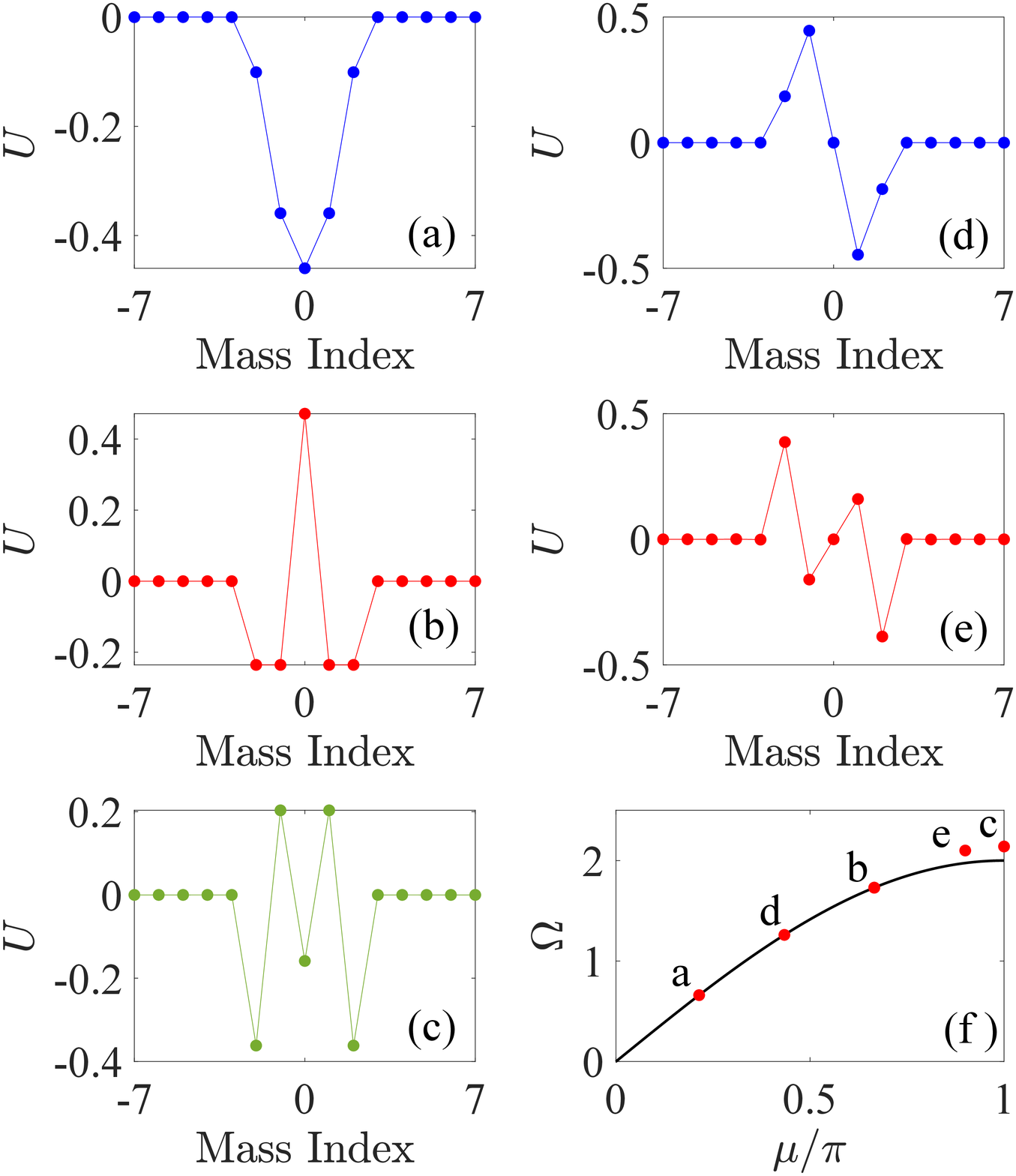}
    \caption{Bound mode shapes in a spring-mass chain: (a)-(c) Even bound mode shapes (d)-(e) Odd bound mode shapes. (f) Dispersion diagram of spring mass system along with the frequencies of bound modes in markers. Modes in (a), (b), and (d) lie in the passband and are thus BICs.}
    \label{fig:boundMode}
\end{figure}

Figures~\ref{fig:boundMode}a-\ref{fig:boundMode}c displays the three bound mode shapes corresponding to these frequencies. The horizontal axis is for location along the $x$-axis of the chain and the vertical axis displays the normalized displacement of the mode shape. In these figures, the displacements of masses only in the top chain are shown. Recall that the modes are symmetric about the $x$ axis and thus the top and bottom chains have identical displacements in each corresponding mass. In addition, these modes are also symmetric about the $y$-axis as expected. 

Similarly, let us determine the mode shapes of odd bound modes. In this case, anti-symmetry about the $y$-axis implies masses $e$ and $e^\prime$ stay at rest. In addition, $u_b=-u_{b^\prime}$ and similar constraints hold for masses $c$, $b^{\prime\prime}$ and $c^{\prime\prime}$. Using these relations and the constraints on masses $d$ and $a$  discussed previously for the even modes, the unknown displacements are $u_b$ and $u_c$. Therefore, the mass and stiffness matrices corresponding to the displacement vector $\bm{u}_o = (u_b, u_c)^T$ are 
\begin{equation*}
    \bm{M}_{o} = \begin{bmatrix}
    1 &0\\
    0 &1 
    \end{bmatrix},  \quad \bm{K}_{o} = \begin{bmatrix}
    4 &-1\\
    -1 &2 
    \end{bmatrix}. 
\end{equation*}
The resulting eigenvalue problem $\bm{K}_o \bm{u}_o = \Omega^2 \bm{M}_o \bm{u}_o$ gives the odd bound mode frequencies 1.26 and 2.10. The corresponding defect masses, given by $\alpha=6/\Omega^2$ are 3.78 and 1.36, respectively. Figures~\ref{fig:boundMode}d and \ref{fig:boundMode}e display the mode shapes at these two frequencies. Again, only the displacement of the masses in the top chain are plotted. As expected, these modes are anti-symmetric about the $y$-axis. 

Finally, recall that a bound mode's frequency must lie in the passband to qualify as a BIC. Let us determine the passband of the infinite spring mass chain without defects. We do a dispersion analysis seeking traveling wave solutions of the form $u(n,\tau) = Ue^{i (\Omega \tau-\mu n)}$. Here $n$ is a mass index and $\mu$ is the non-dimensional wavenumber related to the wavenumber $\kappa$ by $\mu = \kappa l$. Substituting this solution form into the governing equation for mass $n$ gives the frequency wave-number relation $\Omega^2 =  2(1-\cos \mu)$. Figure~\ref{fig:boundMode}f displays this relation in the irreducible Brillouin zone $\mu\in[0,\pi]$, showing a passband when $\Omega\leq 2$ and a stopband at $\Omega >2$. In addition, the figure has markers indicating the frequencies of the various bound modes. We see that the frequencies of modes in Fig.~\ref{fig:boundMode}a,~\ref{fig:boundMode}b, and~\ref{fig:boundMode}d lie in the pass band and they are thus the valid BICs in the considered structure.

\begin{figure}[!t]
    \centering
    \includegraphics[height = 9cm]{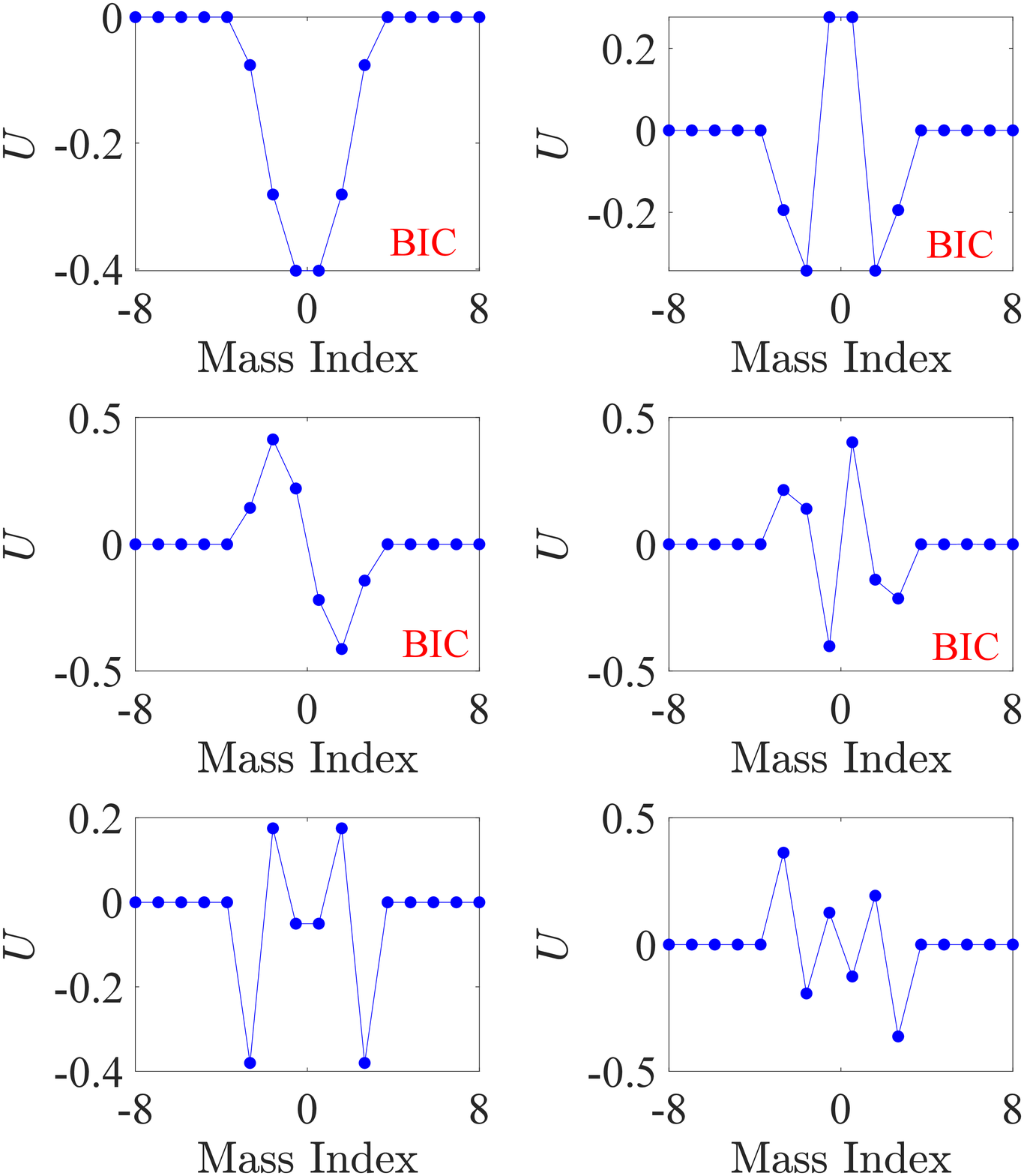}
    \caption{Bound mode shapes in a spring mass chain having 6 masses in the compact region. Frequencies of four bound modes lie in the passband and are thus BICs, as indicated. }
    \label{fig:N6}
\end{figure}

The procedure introduced above can be used to determine a family of bound modes for any size of the compact region, i.e., any number of masses enclosed by the two defects. We performed calculations with various sizes of the compact region and found a family of BICs in each case. Figure~\ref{fig:N6} displays the bound modes in a spring mass chain with 6 masses in the compact region between the defects. Out of these 6 modes, the indicated 4 modes are BICs as their frequency lies in the passband. In general, if we have $2n-1$ (or $2n$) masses between defects in each chain, then there are $n$ even bound modes and $n-1$ (or $n$) odd bound modes. Explicit calculations can be used to determine which of these bound modes are BICs, i.e., those with frequencies in the passband.

\begin{figure}[!b]
    \centering
    \includegraphics[height = 8.75cm]{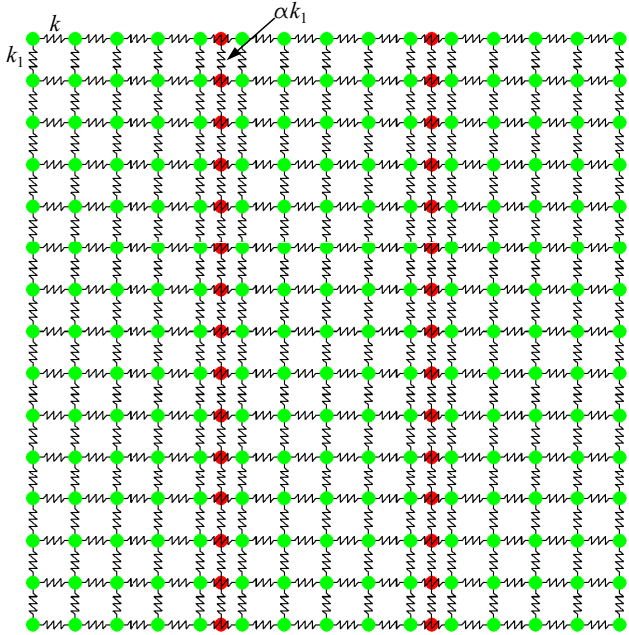}
    \caption{Top view of the waveguide. It consists of an array of the coupled chains in Fig.~\ref{fig:chain}b, with the masses in adjacent chains connected with springs of stiffness $k_1$. The corresponding defect masses are connected by springs of stiffness $\alpha k_1$.}
    \label{fig:waveguide_top}
\end{figure}
\subsection{BIC propagation through waveguide}

Let us demonstrate how to achieve propagation of a BIC mode along a waveguide. Figure~\ref{fig:waveguide_top} displays a schematic of the proposed waveguide concept. It consists of an array of bi-layered spring mass chains coupled with two defects, identical to the chains in Fig.\ref{fig:chain}b. The masses (green nodes) in each chain are coupled with the corresponding masses in adjacent chains by springs of stiffness $k_1$. In addition, the defect masses (red nodes) in adjacent chains are connected by springs of stiffness $\alpha k_1$. Hence we have a spring mass lattice that is periodic along $z$ and whose unit cell is the spring mass chain in Fig.~\ref{fig:chain}b.

\begin{figure}[!t]
        \centering
           
              \includegraphics[height=3.75cm]{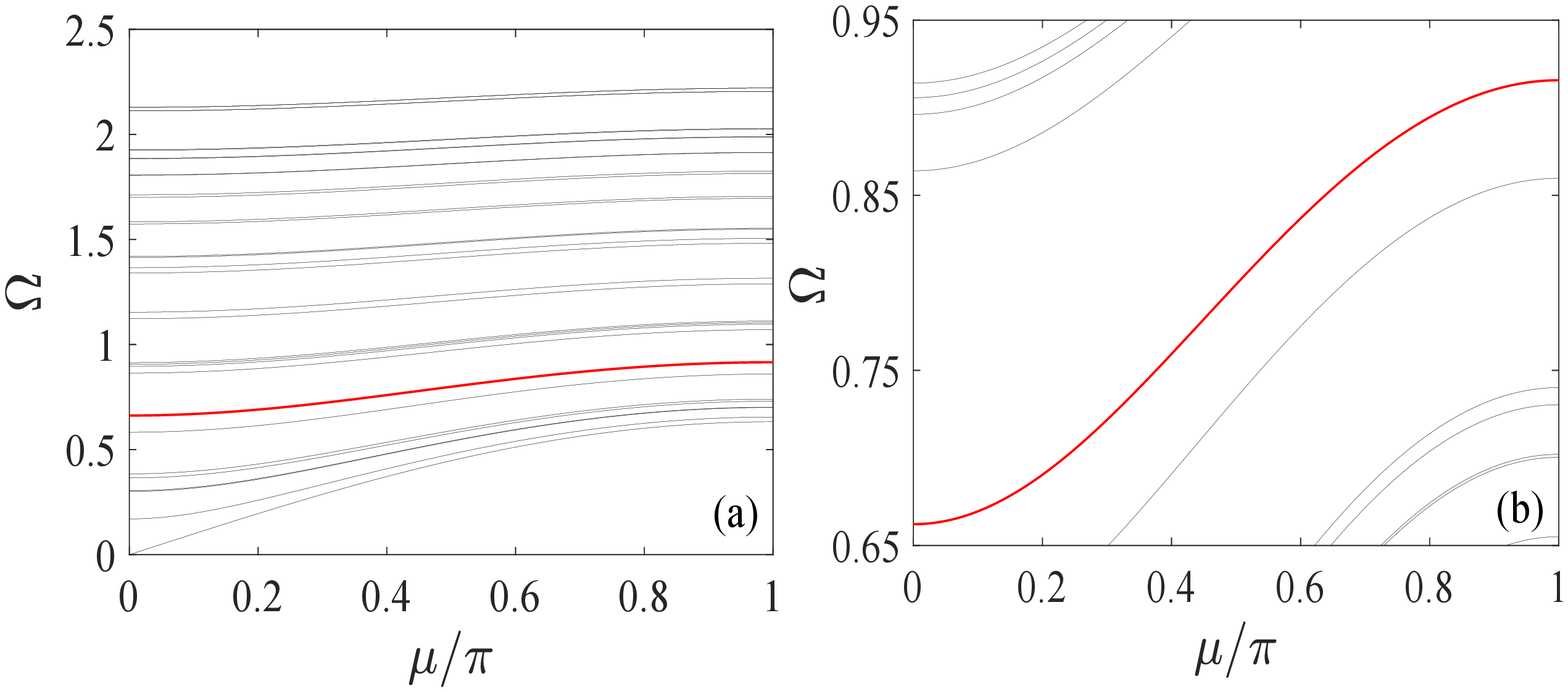}
              \label{fig:guide_disp}

           \caption{(a) Dispersion diagram of the waveguide whose unit cell is the chain in Fig.~\ref{fig:chain}b. The red branch has BIC modes at all wavenumbers $\mu$, indicating the possibility of BIC wave transport. (b) Zoomed in view of BIC mode branch.}
           \label{fig:guide_disp}

\end{figure}
\begin{figure}[!t]
    \centering
    \includegraphics[height=10.5cm]{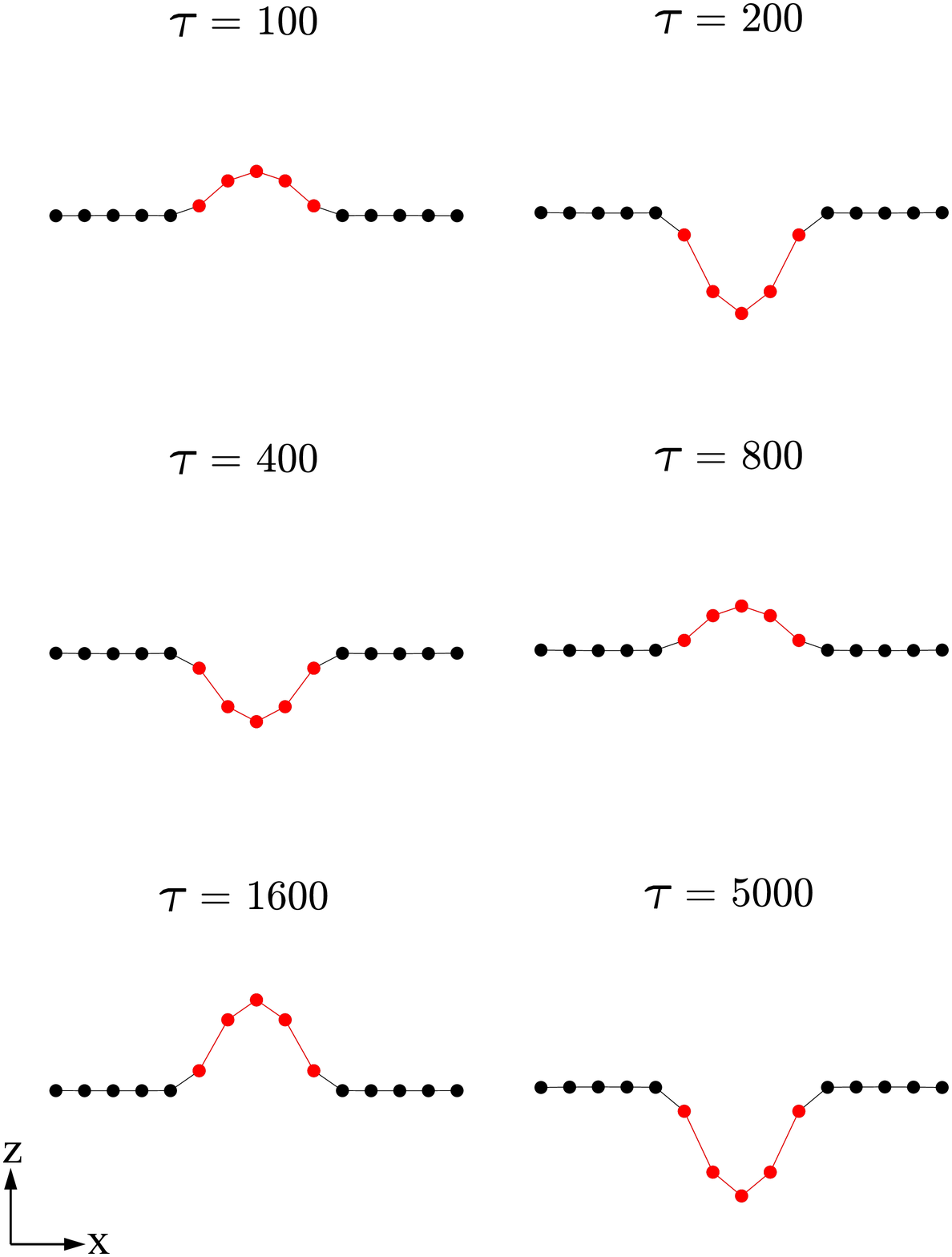}
    \caption{Snapshots of transient simulation showing the displacement field along the top chain at various time instants $\tau$ . After initial transients, the vibration is fully confined in a compact region. }
    \label{fig:trans_BIC}
\end{figure}

\begin{figure*}[!t]
    \centering
    \includegraphics[height = 5cm]{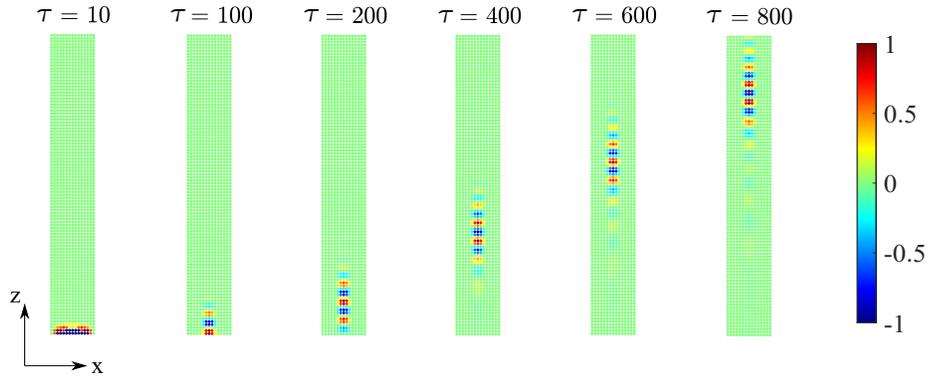}
    \caption{Snapshots of transient simulation showing contours of the normalized displacement field of top layer at various times $\tau$. The BIC mode remains confined and propagates along the waveguide. }
    \label{fig:BIC_transport}
\end{figure*}
As discussed earlier, each unit cell supports BICs for specific value of defect masses. When connected with each other, the lattice has a channel bounded by the two defects in each unit cell. We will show below how this lattice supports a BIC mode propagating along the channel. First, let us discuss the rationale for scaling the stiffness of springs connecting defect masses by $\alpha$. To propagate a BIC through the waveguide, we need equal and opposite displacement of masses $b$ and $d$ in every unit cell at all times. This requirement $u_b = -u_d$ implies $\ddot{u}_b = -\ddot{u}_d$, or equivalently, the ratio of inertial forces is $m_d \ddot{u}_d / m_b\ddot{u}_d = -\alpha$. To induce equal magnitude of acceleration on masses $b$ and $d$ from forces due to adjacent unit cells, the ratio of forces on mass $d$ from springs connecting it to defects in adjacent unit cells to corresponding forces on mass $b$ from adjacent unit cells is also $\alpha$. Recalling the requirement of $u_b = -u_d$ in each unit cell, the defects are connected by spring of stiffness $\alpha k_1$ to satisfy this force ratio condition.

Let us now analyze the wave propagation characteristics along the waveguide by performing a dispersion analysis. The governing equation for the masses in unit cell $n$ may be written in matrix form as 
$\bm{M}\ddot{\bm{u}}_n + \bm{Ku}_n + \bm{K}_1(\bm{u}_n - \bm{u}_{n-1}) + \bm{K}_1(\bm{u}_n - \bm{u}_{n+1}) = \bm{0}$.
Here $\bm{u}_n$ is a vector having the displacements of all masses in unit cell $n$. $\bm{M}$ is the mass matrix, the stiffness matrix $\bm{K}$ contains the  terms due to springs in unit cell $n$, while the stiffness matrix  $\bm{K_1}$ is for the springs that are connected to adjacent unit cells. Again, we seek traveling wave solutions of the form $\bm{u}_n(\tau) = \bm{u} e^{i(\Omega \tau - \mu n)}$. Here $\Omega$ and $\mu$ are the non-dimensional frequency and wavenumber, respectively. Substituting this solution form into the governing equation leads to the eigenvalue problem $    \left( \bm{K} + 2(1 -\cos \mu)\bm{K}_1 \right)\bm{u} = \Omega^2\bm{M}\bm{u}$. Its solution gives the dispersion curves and the number of curves is equal to the number of degrees of freedom in the unit cell. Figure~\ref{fig:guide_disp}a displays these curves for $\alpha = 13.68$ and $k_1 = 0.1 k$. This value of $\alpha$ is chosen as it leads to the BIC mode in Fig.~\ref{fig:boundMode}a in a single unit cell. Note that the group velocity of waves increases with $k_1$. The frequencies at $\mu=0$ are the same as the natural frequencies of an isolated unit cell. The red branch in the dispersion diagram has BIC modes at all wavenumbers. This branch has non-zero group velocity for $\mu \in (0,1)$, indicating the potential of this lattice to propagate BIC modes along the channel bounded by the two defect masses.  

\section{Numerical simulations of a BIC mode and a waveguide}\label{sec:numer}

Having analytically predicted the existence of BICs in the considered spring mass chains, let us now verify using numerical simulations if they remain bound in a chain and if they propagate along a confined channel in a lattice. For these calculations, we set the defect mass  $\alpha =  13.68$ that supports the BIC shown in Fig.~\ref{fig:boundMode}a at frequency $\Omega=0.66$.  A 4th order Runge-Kutta solver is used to conduct transient simulations of the set of governing equations with time step $\tau = 0.07 $.

Let us first investigate the dynamic response of the coupled spring mass chains in Fig.~\ref{fig:chain}b. We subject the center masses $e$ and $e^\prime$ in both chains to a half cycle of windowed tone burst force excitation of the form  $f = \sin(\Omega \tau/N)\sin(\Omega \tau)$. Here $\Omega =  0.66$ and $N =  50$, ensuring that we excite a narrow band of frequencies centered at the BIC frequency. The $4$ masses at the left and right boundary are fixed, while the mass adjacent to the fixed mass at each boundary is subject to critical damping $c=2$. This damping, applied to 4 masses in the structure, damps out other modes which are induced in the structure due to the excitation of a finite bandwidth. 

Figure~\ref{fig:trans_BIC} displays the displacement of the top spring mass chain at various times $\tau$. The half cycle excitation lasts until about $\tau=237$. The BIC arises in the chain at short times, $\tau = 100$. Notably, even after a long time $\tau = 5000$, the masses in the compact region (marked with red nodes in Fig.~\ref{fig:trans_BIC}) vibrate with significant amplitude and the energy is confined within the compact region. This simulation verifies the existence of BICs and illustrates how they can be excited by subjecting a single mass in each chain to an external force. 

Next, let us subject the lattice to a dynamic excitation and analyze if a transversely bound mode corresponding to a BIC can propagate along the channel bound by the defects. We consider a finite structure having $100$ unit cells along the $z$-direction, with each unit cell being identical to the coupled chain in Fig.~\ref{fig:chain}b. We set the spring stiffness $k_1$ between unit cells to $0.1k$ and defect mass $\alpha=13.68$, which supports the BIC in  Fig.~\ref{fig:boundMode}a. 
Again, the masses at the left and the right boundaries are fixed and critical damping is applied to the masses adjacent to the fixed masses. A half cycle of windowed tone burst force excitation of the form $f_i = u_i\sin(\Omega \tau/N)\sin(\Omega \tau)$ is applied to 3 masses in each layer: $c$, $e$ and $c^\prime$ in the top layer and the corresponding ones in the bottom layer on the first unit cell from the lower end. The amplitude of force $u_i$ applied to each mass is proportional to its displacement in the BIC mode shape (in Fig.~\ref{fig:boundMode}a). Here $N=50$ and the forcing frequency is set to $\Omega = 0.73$ that corresponds to wavenumber $\mu=\pi/3$ in the BIC dispersion branch. Note from the dispersion diagram (Fig.~\ref{fig:guide_disp}a) that there are multiple modes  that exist at this frequency. We apply the external force to multiple masses in order to predominantly excite the BIC mode shape. 

Figure~\ref{fig:BIC_transport} displays contours of the  displacement field in the top layer at various times $\tau$. At small times, the waves propagate outside the compact region. As discussed above, there are multiple bulk modes in addition to the BIC at the excitation frequency $\Omega$ and these bulk modes lead to vibrations outside the compact region. As time progresses, these vibrations are damped out by damping at the ends of the chains and we see a wave propagating in the channel bound by the defects. This propagating wave corresponds to mode at $\mu=\pi/3$ on the BIC branch in Fig.~\ref{fig:guide_disp}a. Note that the wave is confined even at long propagation distances, verifying the proposed concept of a waveguide for waves that are BIC in the transverse direction (along $x$). More importantly, it shows the possibility of guiding waves along a channel at frequencies that lie in the passband of the structure, thus eliminating the need for bandgaps.

\section{Conclusions}\label{sec:conc}

The first part of this work analytically predicts the existence of a family of BICs in a bi-layered spring mass chains coupled by two defects. Bound mode shapes and their frequencies are determined by exploiting the reflection symmetry in the compact region. The number of bound modes is equal to the number of masses in a chain between the defects and each bound mode requires a distinct defect mass. The analytical predictions are verified by numerical simulations of transient response, which confirm that a BIC arises in the spring mass chain and the mode does not leak energy to surrounding for a long time after force is removed. 

The second part presents a waveguide made of an array of BIC supporting coupled spring mass chains. The waveguide supports the propagation of a transversely bound wave along a channel between two defects. Dispersion analysis along the propagation direction reveals the presence of a branch having transverse BICs. This branch lies in the passband of the structure. Numerical simulations of a finite lattice excited at a frequency lying in the BIC branch show the propagation of a bound wave along a channel in the waveguide without any energy leakage. 

The framework we present here allows to achieve  BICs in a compact region of arbitrary size and at an arbitrary location in a long chain. Due to their zero energy leakage property, BIC based resonators can achieve extremely high quality factors, in principle infinity, in contrast to conventional resonators. In addition, BIC based waveguides allow for wave propagation along a channel in structures without requiring any bandgaps. Bandgaps are typically hard to achieve and impose stringent geometric constraints. The BIC based waveguides presented here may be a strategy to avoid the requirement of bandgaps. Finally, the presented design is simple and independent of material properties or length scales. These concepts can be extended to achieve BICs in a variety of architected structures or meta-structures.

\section*{Acknowledgements}
This work was supported by the U.S. National Science Foundation under Award No. 2027455.

\bibliography{aipsamp}

\end{document}